\documentclass[conference]{IEEEtran}
\usepackage{graphicx}         
\usepackage{graphicx}
\usepackage {subfigure}     
\usepackage{epstopdf}
\usepackage{subfloat}

\usepackage[table]{xcolor}
\usepackage{color}
\usepackage[export]{adjustbox}

\renewcommand{\figurename}{Figure}
\begin{document}
%
\title{FPGA Implementation of SIMON-128 Cryptographic Algorithm Using Artix-7}


\author{
\IEEEauthorblockN{Ridha Ghayoula$^{1,2}$, Jaouhar Fattahi$^3$,  Amor Smida$^4$, Issam El Gmati$^5$, Emil Pricop$^6$ and Marwa Ziadia$^3$} 
\IEEEauthorblockA{$^1$Unit of Research in High Frequency Electronic Circuits and Systems, University of Tunis El Manar- 2092, Tunis, Tunisia.}

\IEEEauthorblockA{$^2$Department of Electrical and Computer Engineering, Laval University, Quebec, Canada.}
\IEEEauthorblockA{$^3$Department of Computer Science and Software Engineering, Laval University, Quebec, Canada.}
\IEEEauthorblockA{$^4$Department of Medical Equipment Technology, College of Applied Medical Sciences, 
Majmaah University,\\ Majmaah, 11952, KSA.}
\IEEEauthorblockA{$^5$College of Engineering at al Gunfudha Umm Al Qura University, KSA.}
\IEEEauthorblockA{$^6$Automatic Control, Computers and Electronics Department. Petroleum-Gas University of Ploiesti, Romania.} 
}

\maketitle

\begin{abstract}

FPGA is a hardware architecture based on a matrix of programmable and configurable logic circuits thanks to which a large number of functionalities inside the device can be modified using a hardware description language. These functionalities must often be secured especially when the context is sensitive (military, banking, medical, legal, etc.). In this paper, we put forward an efficient implementation of SIMON's block cipher algorithm using Xilinx Vivado 2018.2. The proposed design is analyzed through simulation on Xilinx Artix-7.  A prototype of our design is implemented using the xc7a35tcsg324-1 FPGA chip. Performance and results are discussed.  

\end{abstract}

\vspace{0.2cm}



%
\IEEEpeerreviewmaketitle

\section{Introduction} 

Cryptography \cite{art18} has been used for thousands of years. Nowadays, it is more and more present in our daily life. Contemporary cryptography is mainly interested in, but not limited to, the six following properties:
\begin{enumerate}

\item Secrecy (or confidentiality): the property that ensures that secret or sensitive information is not discovered by an unauthorized party;

\item Authentication: the property that ensures that a stakeholder or service requester is who they claim to be, by presenting something they have, something they are, or something they know. 

\item Integrity: the property that ensures that the information has not been modified in a malicious, accidental, or intentional way by a third party;

\item Authenticity: authentication and integrity usually results in authenticity, which is the guarantee that the information is authentic and comes really from its purported source as it is sent;

\item Non-repudiation: the property that ensures that a stakeholder cannot deny his action, or his partial participation in an action. This results in the fact that one can always establish irrefutable proof of an stakeholder's action;

\item Availability \cite{7861110}: the property that ensures that the information or service has not ceased to exist in a malicious way.

\end{enumerate}

The ability to keep an encrypted message secret is based not on the encryption algorithm but on a secret piece of information called a key that must be used with the algorithm to produce the encrypted message. The size of the key, expressed in number of bits, is a key element and plays a crucial role in the security of the cryptographic algorithm. Depending on whether the key used for encryption and decryption is the same or not, we speak of a symmetric or asymmetric cryptographic system. Symmetric cryptography, also known as secret key cryptography, uses a unique key to encrypt and decrypt data. This key must be shared with the recipient. The advantage of symmetric cryptography is that it is easy to implement. Its disadvantage is that the secret key must be shared with the recipient, which adds a key management burden. Unlike symmetric cryptography, asymmetric cryptography requires two keys for its operation: first, a so-called public key that must be made public to recipients; second, a private key that must be kept secret. The public key and the private key are two totally different things, nevertheless, they are linked by mathematical bonds. The advantage of asymmetric cryptography is that one does not manage the security of key sharing, but its disadvantage is that it takes a lot of time. Also, encrypted messages are much larger than those encrypted using symmetric keys. \\

Cryptographic protocols, using symmetric or asymmetric cryptographic algorithms, \cite{art17,art17-2,art17-3,art17-4} are rules of exchange between network points whose role is precisely to secure communications. They are used for example in e-commerce, when a customer enters his credit card number to pay for a purchase. But they are also used in a multitude of other situations, such as when connecting to a computer in a secure manner, when sending e-mails if one wishes to prevent an eavesdropper from reading them, or when checking one's bank account balance. They are used for any use of the bank card, such as withdrawing money from an ATM or paying in a restaurant. They have also been used for a long time in the decoders of pay TV channels to allow the customer to have access to the channels to which he has subscribed and to prevent him from accessing other channels, while allowing possible changes in the subscription. Nowadays, it is possible to implement a cryptographic algorithm in a software or hardware way. The hardware implementation of a cryprographic algorithm consists of the use of computer hardware (e.g. processors, dedicated chips, etc.) in the data encryption process. In general, this implementation is put integrated in the instruction set of the processor, which means that a part of the processor is dedicated to the cryptographic mission. This also means that a significant increase in speed will be observed. By the same stream of ideas, parallel architectures of modern processors are capable of executing other instructions at the same time. A secure cryptoprocessor is a processor optimized for cryptographic tasks (modular exponentiation, DES encryption, etc.) incorporated with multiple physical security measures, giving it some resistance to tampering. It can be realized in various ways depending on the profile of the use: FPGA, ASIC or microcontroller. The reconfiguration capabilities of FPGAs allow considerable optimization of operations and correction of implementations if necessary. Some encryption algorithms are less suitable than others for modern hardware. DES, for example, is based on permutations between bits, which may not be suitable for some types of hardware. \\


In this paper, we propose a hardware implementation of the SIMON-128 cryptographic algorithm\cite{pap1, pap2,pap3, pap4}. The architecture we are putting forward is designed using Xilinx Vivado 2018.2. It is implemented on the xc7a35tcsg324-1 Artix-7 FPGA board. Artix-7\cite{Artix-71,Artix-72} is a development platform designed around the Xilinx in-situ programmable gate array. It is designed entirely for use as a MicroBlaze softcore processor system. The Artix-7 FPGA is optimized for high-performance logic and offers better  performance as well as more capacity than old designs.\\




The methodology including the SIOMON algorithm description is presented in section II. Design, synthesis,  implementation, and experimental results  are presented in section III. Discussion is made in section IV. Finally, section V makes conclusions.

\section{Methodology}

\begin{figure} [!ht]
\centerline{\includegraphics[width=7.5cm,height=7cm,frame]{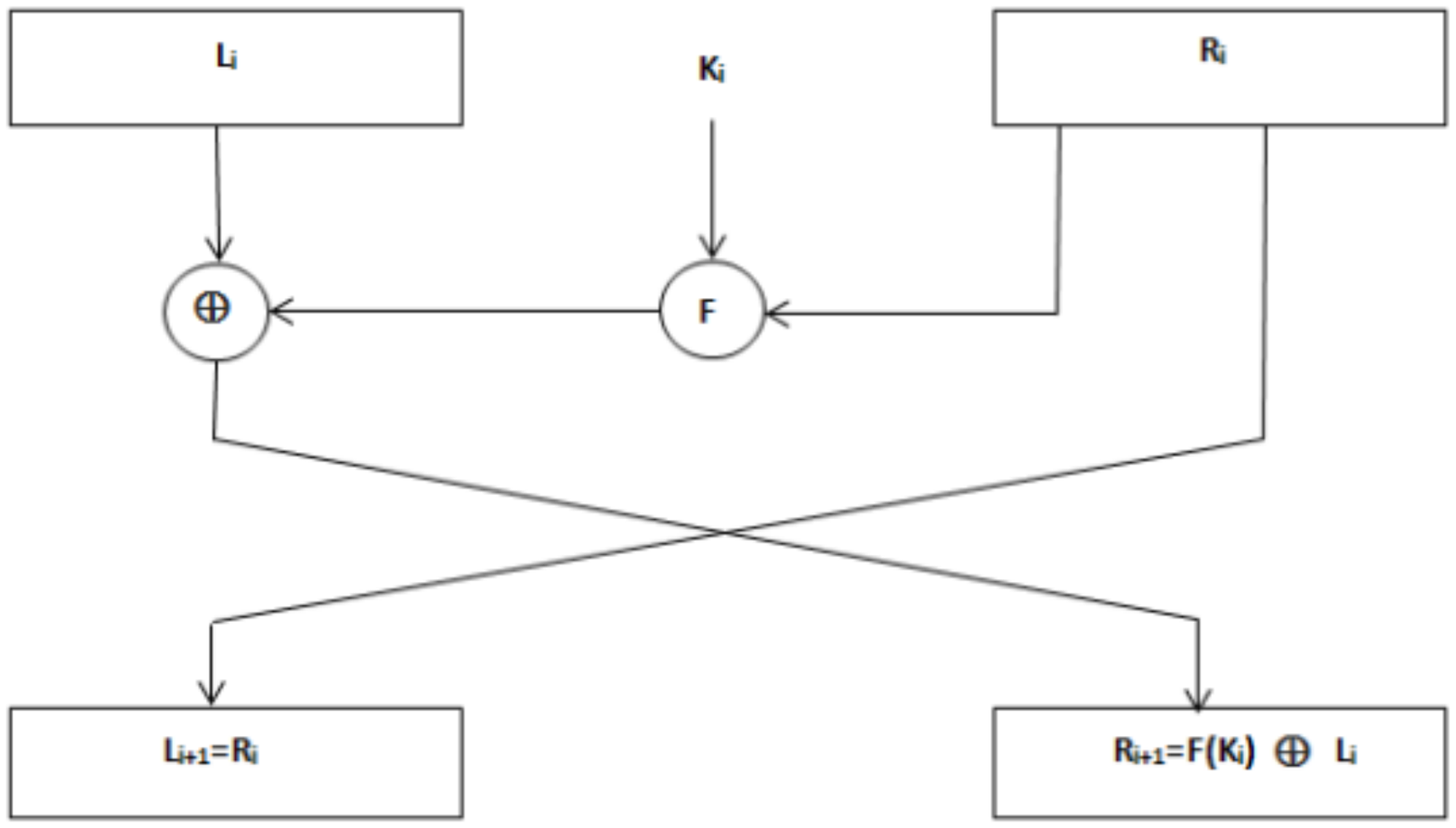}}
\caption{Feistel round.}
\label{figFeistel}
\end{figure}

The algorithm we implement in this paper is the SIMON encryption algorithm. It was proposed by researchers in cryptography at the NSA. One of SIMON's security goals was to keep a reasonable level of security in an environment where power, memory and processors are severely limited. The detailed description of the algorithm is freely available on the web. SIMON is a symmetric block cipher algorithm. The computation scheme used is a Feistel network\cite{feis1,feis2}. Feistel's process \ref{figFeistel} hinges on the idea that repeating judiciously chosen simple operations enough times allows good security. Encryption is a succession of similar steps (called rounds) each using a subkey. This process is characterized by:

\begin{enumerate}
\item An iterative and modular construction;
\item Subkeys are derived from the secret key;
\item The functions used by a lathe must be optimized and are generally simple operations.
\end{enumerate}

These simple operations are generally:
\begin{enumerate}
\item Permutation: the symbols of the plain text are exchanged between them. Permutation adds diffusion;
\item Substitution: a symbol is replaced by another symbol. Substitution adds confusion.
\end{enumerate}

The encryption is decomposed into several rounds. In each round, two blocks are exchanged and one block is combined with a transformed version of the second block and a key. The transformation function is a non-linear bijection while the combination function is usually the exclusive function (XOR). The rotating key is generated from an initial secret key. The key generation mechanism is usually called a key program. This scheme thus provides the two properties of diffusion and confusion necessary in an encryption algorithm. SIMON is an algorithm intended to be physically implemented in highly constrained embedded systems. For this the following design choices have been made \cite{art6, art7, art14}:

\begin{itemize}
\item The size of the block and the key are configurable;
\item The number of laps required for encryption then varies;
\item The nonlinear function used at each turn is very simple;
\item The complexity of decryption grows exponentially with the number of rounds.
\end{itemize}

SIMON respects the symmetry operation regarding the circular shift operation on n-bit words. The key schedule uses a succession of 1-bit round constants devised for the sliding properties and circular shift symmetry.  

Table \ref{tab1} sums up all configurations of SIMON-128.\\ 

\begin{table}[ht]
\caption{SIMON-128 parameters}
\label{tab1}
\begin{center}
\begin{adjustbox}{width=0.45\textwidth}

\begin{tabular}{|l|l|l|l|l|}
\hline
\multicolumn{1}{|c|}{\textbf{Bloc size (bits)}} & \multicolumn{1}{c|}{\textbf{key size}} & \multicolumn{1}{c|}{\textbf{key word (m)}} & \multicolumn{1}{c|}{\textbf{Round constant}} & \multicolumn{1}{c|}{\textbf{Rounds}} \\ 
\hline
\multicolumn{1}{|c|}{} & \multicolumn{1}{c|}{124} & \multicolumn{1}{c|}{2} & \multicolumn{1}{c|}{$z_2$} & \multicolumn{1}{c|}{68} \\ 
\cline{2-5}
\multicolumn{1}{|c|}{128} & \multicolumn{1}{c|}{192} & \multicolumn{1}{c|}{3} & \multicolumn{1}{c|}{$z_3$} & \multicolumn{1}{c|}{69} \\ 
\cline{2-5}
\multicolumn{1}{|c|}{} & \multicolumn{1}{c|}{256} & \multicolumn{1}{c|}{4} & \multicolumn{1}{c|}{$z_4$} & \multicolumn{1}{c|}{72} \\ 
\hline
\end{tabular}
\end{adjustbox}

\end{center}
\end{table}

Let $S^j$ denote a $j$-bit-left circular shift. The key schedule is mathematically described as:

\begin{table}[!ht]
$\left\{ {\begin{array}{*{20}l}
   {c \oplus {z_{ji} }   \oplus k_i  \oplus \left( {i \oplus S^{ - 1} } \right)\left( {S^{ - 3} K_{i + 1} } \right)} & {, \mbox{ with }  m = 2}  \\ $ $\\
   {c \oplus {z_{ji} }  \oplus k_i  \oplus \left( {i \oplus S^{ - 1} } \right)\left( {S^{ - 3} K_{i + 2} } \right)} & {, \mbox{ with }  m = 3}  \\$ $\\
   {c \oplus {z_{ji} }  \oplus k_i  \oplus \left( {i \oplus S^{ - 1} } \right)\left( {S^{ - 3} K_{i + 3}  \oplus K_{i + 1} } \right)} & {, \mbox{ with } m = 4}  \\
\end{array}} \right.$
\end{table}

The key schedule structure can be either balanced or unbalanced. The number of keywords $m$ is used to establish the key expansion structure, yielding a total bit width of $m*n$. The keyword expansion consists of a right shift, an XOR and a constant sequence, $z_x$. The $z_x$ bit operates on the lowest bit of the keyword once every round \cite{art9, art10}.\\

The SIMON-128 encryption is expressed by Equation \ref{eqSimon1}: \\

\begin{equation}
R\left( {l,r,k} \right) = \left( {\left( {S^1 \left( l \right) \& S^8 \left( l \right)} \right) \oplus S^2 \left( l \right) \oplus r \oplus k,l} \right)\\
\label{eqSimon1}
\end{equation}

$ $\\

The SIMON-128 decryption is expressed by Equation \ref{eqSimon2}: \\

\begin{equation}
R^{ - 1} \left( {l,r,k} \right) = \left( {r,\left( {S^1 \left( r \right) \& S^8 \left( r \right)} \right) \oplus S^2 \left( r \right) \oplus l \oplus k} \right)\\
\label{eqSimon2}
\end{equation}
$ $\\

Where $l$ is the left-most word of a block, $r$ the right-most word and $k$ the corresponding round key \cite{art12}.

\section{Implementation and experimental results }

We assume that the block of data to be encrypted as well as the key are presented at the same time. A pulse of the "start" signal indicates that the encryption can start. Once the encryption is finished, the "done" signal goes to one indicating that the value presented on the output "ciphertext" is valid. In addition to these signals, we also have an input for the clock signal "clk" and reset "nrst".
The module is divided into three parts:
\begin{itemize}
\item SIMON$_{dp}$: for encrypting data blocks;
\item SIMON$_{ks}$: for generating the turn key;
\item SIMON$_{ctrl}$: for generating control signals.
\end{itemize}

\begin{figure} [!ht]
\centerline{\includegraphics[width=0.5\textwidth,draft=false]{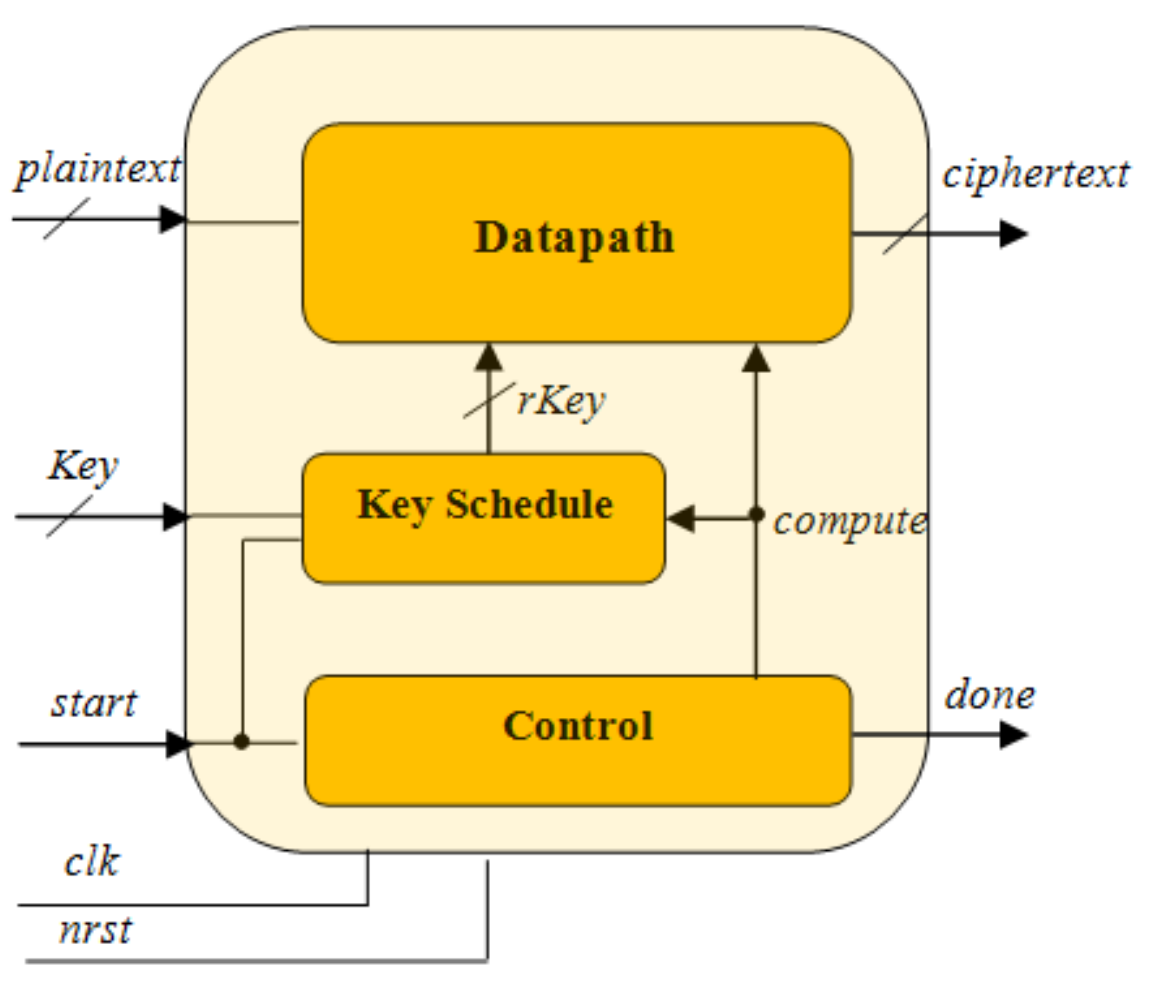}}
\caption{SIMON Design.}
\label{fig2}
\end{figure}

The architecture of the SIMON cipher block consists of a parallel cipher that uses round functions and key generation blocks. 

\begin{figure} [!ht]
\centerline{\includegraphics[width=0.5\textwidth,draft=false]{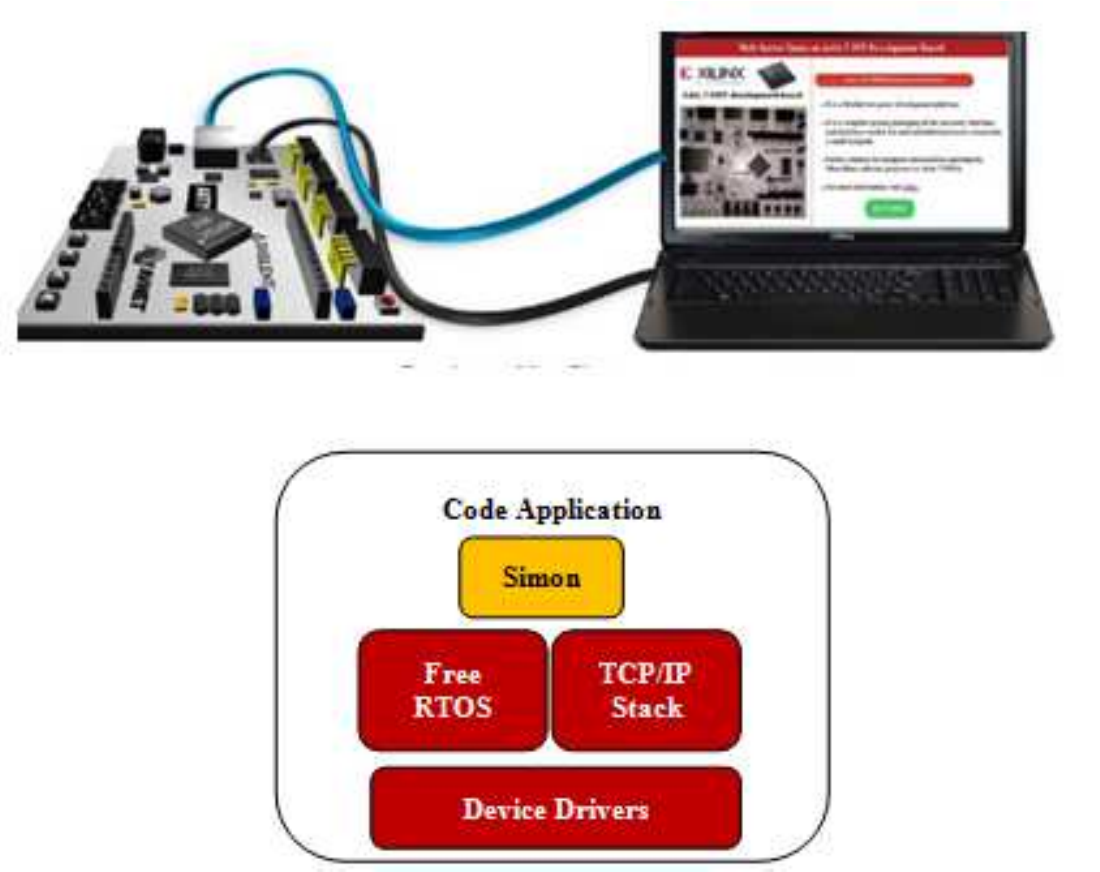}}
\caption{SIMON-Code application using Artix-7.}
\label{fig3}
\end{figure}

The SIMON-128 architecture is implemented on a Xilinx Arty board which is a Artix-7 Pro-based embedded development platform.  The $xc7a35tcsg324-1$  FPGA contains $20800$ LUT, $4600$ Flip Flip and $9600$ LUTRAM modules. We used Xilinx Vivado $2018.2$ Softwares to implement our architecture on the board. All VHDL modules are extensively simulated using Vivado $2018.2$ and synthesized using Xilinx synthesis technologies. Figure \ref{fig3} shows the experimental setup for the SIMON-$128$ architecture. Table \ref{tab31} presents the implementation resources (Post-synthesis and post-implementation).

\begin{table}[!ht]
\caption{Implementation resources (Post-synthesis and post-implementation)}
\begin{center}
\renewcommand{\arraystretch}{1.25}
\small{

\begin{tabular}{|l|l|l|l|}
\hline
\multicolumn{1}{|c|}{\textbf{Resource}} & \multicolumn{1}{c|}{\textbf{Utilization}} & \multicolumn{1}{c|}{\textbf{Available}} & \multicolumn{1}{c|}{\textbf{Utilization $(\%)$}} \\ 
\hline
\multicolumn{1}{|c|}{\textbf{LUT}} & \multicolumn{1}{c|}{45} & \multicolumn{1}{c|}{20800} & \multicolumn{1}{c|}{0.22} \\ 
\hline
\multicolumn{1}{|c|}{\textbf{LUTRAM}} & \multicolumn{1}{c|}{12} & \multicolumn{1}{c|}{9600} & \multicolumn{1}{c|}{0.13} \\ 
\hline
\multicolumn{1}{|c|}{\textbf{FF}} & \multicolumn{1}{c|}{27} & \multicolumn{1}{c|}{4600} & \multicolumn{1}{c|}{0.06} \\ 
\hline
\multicolumn{1}{|c|}{\textbf{IO}} & \multicolumn{1}{c|}{5} & \multicolumn{1}{c|}{210} & \multicolumn{1}{c|}{2.38} \\ 
\hline
\multicolumn{1}{|c|}{\textbf{BUFG}} & \multicolumn{1}{c|}{1} & \multicolumn{1}{c|}{32} & \multicolumn{1}{c|}{3.13} \\ 
\hline
\end{tabular}
}
\label{tab31}
\end{center}
\end{table}

Summary of on-Chip static and dynamic power are shown in  Table \ref{tab4} using $xc7a35tcsg324-1$ device of Artix-7 family. Table \ref{tab4} and Table \ref{tab5} present the thermal and power characteristics of this implementation. 

\begin{table}[!ht]
\caption{On-chip dynamic and static power}
\label{tab4}
\begin{center}


\begin{adjustbox}{width=0.47\textwidth}
\renewcommand{\arraystretch}{1.9}
\begin{tabular}{|l|l|l|l|l|l|}
\hline
\multicolumn{3}{|c|}{\textbf{Dynamic}} & \multicolumn{3}{c|}{\textbf{Static}} \\ 
\hline
\multicolumn{1}{|c|}{} & \multicolumn{1}{c|}{Power (W)} & \multicolumn{1}{c|}{Percentage} & \multicolumn{1}{c|}{} & \multicolumn{1}{c|}{Power (W)} & \multicolumn{1}{c|}{Percentage} \\ 
\hline
\multicolumn{1}{|c|}{Signals} & \multicolumn{1}{c|}{$<$ 0.001} & \multicolumn{1}{c|}{10\%} & \multicolumn{1}{c|}{} & \multicolumn{1}{c|}{} & \multicolumn{1}{c|}{} \\ 
\cline{1-3}
\multicolumn{1}{|c|}{Logic} & \multicolumn{1}{c|}{$<$ 0.001} & \multicolumn{1}{c|}{13\%} & \multicolumn{1}{c|}{PL Static} & \multicolumn{1}{c|}{0.070} & \multicolumn{1}{c|}{97\%} \\ 
\cline{1-3}
\multicolumn{1}{|c|}{I/O} & \multicolumn{1}{c|}{0.001} & \multicolumn{1}{c|}{46\%} & \multicolumn{1}{c|}{} & \multicolumn{1}{c|}{} & \multicolumn{1}{c|}{} \\ 
\cline{1-3}
\multicolumn{1}{|c|}{Clocks} & \multicolumn{1}{c|}{0.001} & \multicolumn{1}{c|}{31\%} & \multicolumn{1}{c|}{} & \multicolumn{1}{c|}{} & \multicolumn{1}{c|}{} \\ 
\hline
\end{tabular}
\end{adjustbox}

\end{center}
\end{table}

\begin{table}[!ht]
\caption{Thermal and Power Characteristics}
\begin{center}
\renewcommand{\arraystretch}{1.4}
\small{

\begin{tabular}{|l|l|}
\hline
\multicolumn{1}{|c|}{\textbf{Power}} & \multicolumn{1}{c|}{} \\ 
\hline
\multicolumn{1}{|c|}{\textbf{Total On-Chip Power}} & \multicolumn{1}{c|}{0.072 w} \\ 
\hline
\multicolumn{1}{|c|}{\textbf{Junction Temperature}} & \multicolumn{1}{c|}{  $25.3^ \circ$} \\ 
\hline
\multicolumn{1}{|c|}{\textbf{Thermal margin}} & \multicolumn{1}{c|}{$59.7^ \circ$ $(12.)$ w} \\ 
\hline
\multicolumn{1}{|c|}{\textbf{Effective JA}} & \multicolumn{1}{c|}{$4.8^ {\circ}$ c/w } \\ 
\hline
\multicolumn{1}{|c|}{\textbf{Power Supplied to off-hip devices}} & \multicolumn{1}{c|}{0 w} \\ 
\hline
\multicolumn{1}{|c|}{\textbf{Confidence level}} & \multicolumn{1}{c|}{Medium} \\ 
\hline
\end{tabular}
}
\label{tab5}
\end{center}
\end{table}

\begin{figure*} [!ht]
\centerline{\includegraphics[width=1.\textwidth,draft=false]{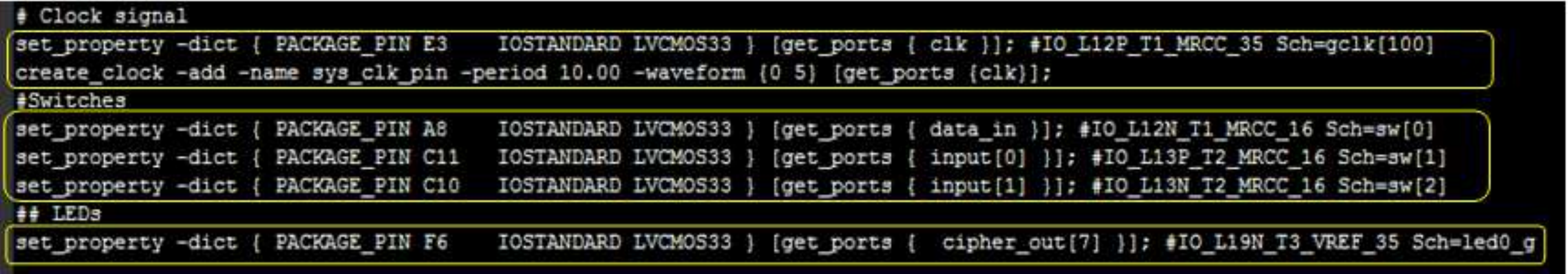}}
\caption{SIMON-128 XDC file.}
\label{fig4}
\end{figure*}


\section{Discussion} 

Table \ref{tabresult} gives  the performance of our implementation and compares it with the results obtained with Zynq-7000 and Virtex-7 presented in \cite{art16}. Our implementation presents significant improvement over both of them regarding all metrics. We note a gain in power consumption of 69.87\% compared to Zynq-7000 and 70.56\% compared to Virtex-7, a gain in delay of 26.69\% compared to Zynq-7000 and 8.95\% compared to Virtex-7. SIMON Artix-7 also uses less area of lookup table (e.g. 45 LUT) compared to both Zynq-7000 and Virtex-7 (73 LUT). Artix-7 confirms its reputation of being quicker and low-cost than other models. 

\begin{table}[!ht]

\caption{Performance and Comparison of our implementation with Zynq-7000 and SIMON Virtex-7}
\label{tabresult}
\begin{center}
\begin{adjustbox}{width=0.5\textwidth}
\renewcommand{\arraystretch}{3}
\begin{tabular}{|c|c|c|c|}

\hline
\textbf{Metric}     & \textbf{\begin{tabular}[c]{@{}c@{}}SIMON Zynq-7000\\ {\cite{art16}}\end{tabular}} & \textbf{\begin{tabular}[c]{@{}c@{}}SIMON Virtex-7\\ {\cite{art16}}\end{tabular}} & \textbf{SIMON Artix-7} \\ \hline
\textbf{Power (mW)} & 239                                                                         & 248                                                                        & 72                     \\ \hline
\textbf{Delay (ns)} & 5.448                                                                       & 4.415                                                                      & 4.020                  \\ \hline
\textbf{Area (LUT)} & 73                                                                          & 73                                                                         & 45                     \\ \hline
\end{tabular}
\end{adjustbox}
\end{center}

\end{table}


Other recent work comparable to our present implementation is worthy of mention. For instance, Rashidi in \cite{Rashidi19} presented an ASIC implementation of  several sizes of the SIMON algorithm using Sklansky adder. Encouraging results were observed regarding critical path delay. In the same vein, Sheikhpour et al. in \cite{Sheikhpour} presented a flexible implementation of SIOMON with various key sizes, which has capabilities for various kinds of attacks. In \cite{Abed}, Abed et al. presented several types of SIMON implementations (pipelined, scalar, etc.) and showed a comparison between these families in terms of throughput and drew up an implementation guideline depending on the technological need.\\

Despite the good performance of our implementation, it requires very sharp knowledge of the hardware and tools and the design is sometimes tedious to set up.

\section{CONCLUSION}

In this paper, we have presented an implementation of SIMON-128 algorithm   using Artix-7of with low-cost FPGA platform. The Feistel feature of SIMON is chosen to reduce the hardware impact of encryption without sacrificing software performance. Such a structure has the advantage that the encryption and decryption operations are very similar, which is enough to reverse the operation of the key manager to obtain the decryption operation. We used circular offsets (just hardware cabling) and bit-to-bit operations. The implementation led to a good performance that we discussed in this paper compared to other implementations in the state of the art.

\clearpage

\bibliographystyle{IEEEtran}
\bibliography{biblioart}

\begin{thebibliography}{10}
\providecommand{\url}[1]{#1}
\csname url@samestyle\endcsname
\providecommand{\newblock}{\relax}
\providecommand{\bibinfo}[2]{#2}
\providecommand{\BIBentrySTDinterwordspacing}{\spaceskip=0pt\relax}
\providecommand{\BIBentryALTinterwordstretchfactor}{4}
\providecommand{\BIBentryALTinterwordspacing}{\spaceskip=\fontdimen2\font plus
\BIBentryALTinterwordstretchfactor\fontdimen3\font minus
  \fontdimen4\font\relax}
\providecommand{\BIBforeignlanguage}[2]{{%
\expandafter\ifx\csname l@#1\endcsname\relax
\typeout{** WARNING: IEEEtran.bst: No hyphenation pattern has been}%
\typeout{** loaded for the language `#1'. Using the pattern for}%
\typeout{** the default language instead.}%
\else
\language=\csname l@#1\endcsname
\fi
#2}}
\providecommand{\BIBdecl}{\relax}
\BIBdecl

\bibitem{art18}
H.~C.~v. Tilborg, \emph{Encyclopedia of Cryptography and Security}.\hskip 1em
  plus 0.5em minus 0.4em\relax Berlin, Heidelberg: Springer-Verlag, 2005.

\bibitem{7861110}
E.~Pricop, S.~F. Mihalache, N.~Paraschiv, J.~Fattahi, and F.~Zamfir,
  ``Considerations regarding security issues impact on systems availability,''
  in \emph{2016 8th International Conference on Electronics, Computers and
  Artificial Intelligence (ECAI)}, 2016, pp. 1--6.

\bibitem{art17}
J.~Fattahi, ``Analyse des protocoles cryptographiques par les fonctions
  témoins,'' Ph.D. dissertation, Laval University, Quebec, Canada, 2 2016.

\bibitem{art17-2}
\BIBentryALTinterwordspacing
J.~Fattahi, M.~Mejri, M.~Ziadia, E.~Ghayoula, O.~Samoud, and E.~Pricop,
  ``Cryptographic protocol for multipart missions involving two independent and
  distributed decision levels in a military context,'' in \emph{2017 {IEEE}
  International Conference on Systems, Man, and Cybernetics, {SMC} 2017, Banff,
  AB, Canada, October 5-8, 2017}.\hskip 1em plus 0.5em minus 0.4em\relax
  {IEEE}, 2017, pp. 1127--1132. [Online]. Available:
  \url{https://doi.org/10.1109/SMC.2017.8122763}
\BIBentrySTDinterwordspacing

\bibitem{art17-3}
\BIBentryALTinterwordspacing
J.~Fattahi, M.~Mejri, M.~Ziadia, T.~Omrani, and E.~Pricop, ``Witness-functions
  versus interpretation-functions for secrecy in cryptographic protocols: What
  to choose?'' in \emph{2017 {IEEE} International Conference on Systems, Man,
  and Cybernetics, {SMC} 2017, Banff, AB, Canada, October 5-8, 2017}.\hskip 1em
  plus 0.5em minus 0.4em\relax {IEEE}, 2017, pp. 2649--2654. [Online].
  Available: \url{https://doi.org/10.1109/SMC.2017.8123025}
\BIBentrySTDinterwordspacing

\bibitem{art17-4}
\BIBentryALTinterwordspacing
J.~Fattahi, M.~Mejri, R.~Ghayoula, and E.~Pricop, ``Formal reasoning on
  authentication in security protocols,'' in \emph{2016 {IEEE} International
  Conference on Systems, Man, and Cybernetics, {SMC} 2016, Budapest, Hungary,
  October 9-12, 2016}.\hskip 1em plus 0.5em minus 0.4em\relax {IEEE}, 2016, pp.
  282--289. [Online]. Available: \url{https://doi.org/10.1109/SMC.2016.7844255}
\BIBentrySTDinterwordspacing

\bibitem{pap1}
\BIBentryALTinterwordspacing
A.~S. Omar and O.~Basir, ``{SIMON} 32/64 and 64/128 block cipher: Study of
  cross correlation and linear span attack immunity,'' in \emph{28th {IEEE}
  Annual International Symposium on Personal, Indoor, and Mobile Radio
  Communications, {PIMRC} 2017, Montreal, QC, Canada, October 8-13,
  2017}.\hskip 1em plus 0.5em minus 0.4em\relax {IEEE}, 2017, pp. 1--6.
  [Online]. Available: \url{https://doi.org/10.1109/PIMRC.2017.8292209}
\BIBentrySTDinterwordspacing

\bibitem{pap2}
\BIBentryALTinterwordspacing
D.~Le, S.~L. Yeo, and K.~Khoo, ``Algebraic differential fault analysis on
  {SIMON} block cipher,'' \emph{{IACR} Cryptol. ePrint Arch.}, p. 436, 2021.
  [Online]. Available: \url{https://eprint.iacr.org/2021/436}
\BIBentrySTDinterwordspacing

\bibitem{pap3}
\BIBentryALTinterwordspacing
S.~M. Dehnavi, ``Further observations on {SIMON} and {SPECK} block cipher
  families,'' \emph{Cryptogr.}, vol.~3, no.~1, p.~1, 2019. [Online]. Available:
  \url{https://doi.org/10.3390/cryptography3010001}
\BIBentrySTDinterwordspacing

\bibitem{pap4}
\BIBentryALTinterwordspacing
R.~Beaulieu, D.~Shors, J.~Smith, S.~Treatman{-}Clark, B.~Weeks, and L.~Wingers,
  ``The {SIMON} and {SPECK} lightweight block ciphers,'' in \emph{Proceedings
  of the 52nd Annual Design Automation Conference, San Francisco, CA, USA, June
  7-11, 2015}.\hskip 1em plus 0.5em minus 0.4em\relax {ACM}, 2015, pp.
  175:1--175:6. [Online]. Available:
  \url{https://doi.org/10.1145/2744769.2747946}
\BIBentrySTDinterwordspacing

\bibitem{Artix-71}
\BIBentryALTinterwordspacing
C.~Fibich, M.~Horauer, and R.~Obermaisser, ``Device- and temperature dependency
  of systematic fault injection results in artix-7 and ice40 fpgas,'' in
  \emph{Design, Automation {\&} Test in Europe Conference {\&} Exhibition,
  {DATE} 2021, Grenoble, France, February 1-5, 2021}.\hskip 1em plus 0.5em
  minus 0.4em\relax {IEEE}, 2021, pp. 1600--1605. [Online]. Available:
  \url{https://doi.org/10.23919/DATE51398.2021.9474161}
\BIBentrySTDinterwordspacing

\bibitem{Artix-72}
\BIBentryALTinterwordspacing
J.~Wang, C.~Feng, W.~Dong, Z.~Shen, and S.~Liu, ``A high precision
  time-to-digital converter based on multi-chain interpolation with a low cost
  artix-7 {FPGA},'' in \emph{7th International Conference on Event-Based
  Control, Communication, and Signal Processing, {EBCCSP} 2021, Krakow, Poland,
  June 22-25, 2021}.\hskip 1em plus 0.5em minus 0.4em\relax {IEEE}, 2021, pp.
  1--5. [Online]. Available:
  \url{https://doi.org/10.1109/EBCCSP53293.2021.9502368}
\BIBentrySTDinterwordspacing

\bibitem{feis1}
\BIBentryALTinterwordspacing
S.~Chen, Y.~Fan, L.~Sun, Y.~Fu, H.~Zhou, Y.~Li, M.~Wang, W.~Wang, and C.~Guo,
  ``{SAND:} an {AND-RX} feistel lightweight block cipher supporting s-box-based
  security evaluations,'' \emph{Des. Codes Cryptogr.}, vol.~90, no.~1, pp.
  155--198, 2022. [Online]. Available:
  \url{https://doi.org/10.1007/s10623-021-00970-9}
\BIBentrySTDinterwordspacing

\bibitem{feis2}
\BIBentryALTinterwordspacing
C.~Guo and G.~Zhang, ``Beyond-birthday security for permutation-based feistel
  networks,'' \emph{Des. Codes Cryptogr.}, vol.~89, no.~3, pp. 407--440, 2021.
  [Online]. Available: \url{https://doi.org/10.1007/s10623-020-00820-0}
\BIBentrySTDinterwordspacing

\bibitem{art6}
\BIBentryALTinterwordspacing
R.~Beaulieu, D.~Shors, J.~Smith, S.~Treatman{-}Clark, B.~Weeks, and L.~Wingers,
  ``The {SIMON} and {SPECK} families of lightweight block ciphers,''
  \emph{{IACR} Cryptol. ePrint Arch.}, p. 404, 2013. [Online]. Available:
  \url{http://eprint.iacr.org/2013/404}
\BIBentrySTDinterwordspacing

\bibitem{art7}
\BIBentryALTinterwordspacing
R.~Beaulieu, D.~Shors, J.~Smith, S.~Treatman-Clark, B.~Weeks, and L.~Wingers,
  ``The simon and speck lightweight block ciphers,'' in \emph{Proceedings of
  the 52nd Annual Design Automation Conference}, ser. DAC '15.\hskip 1em plus
  0.5em minus 0.4em\relax New York, NY, USA: Association for Computing
  Machinery, 2015. [Online]. Available:
  \url{https://doi.org/10.1145/2744769.2747946}
\BIBentrySTDinterwordspacing

\bibitem{art14}
\BIBentryALTinterwordspacing
L.~M.~A. Qassem, T.~Stouraitis, E.~Damiani, and I.~M. Elfadel, ``Fpgaaas: {A}
  survey of infrastructures and systems,'' \emph{{IEEE} Trans. Serv. Comput.},
  vol.~15, no.~2, pp. 1143--1156, 2022. [Online]. Available:
  \url{https://doi.org/10.1109/TSC.2020.2976012}
\BIBentrySTDinterwordspacing

\bibitem{art9}
\BIBentryALTinterwordspacing
A.~Shahverdi, M.~Taha, and T.~Eisenbarth, ``Lightweight side channel
  resistance: Threshold implementations of simon,'' \emph{{IEEE} Trans.
  Computers}, vol.~66, no.~4, pp. 661--671, 2017. [Online]. Available:
  \url{https://doi.org/10.1109/TC.2016.2614504}
\BIBentrySTDinterwordspacing

\bibitem{art10}
\BIBentryALTinterwordspacing
A.~Aysu, E.~Gulcan, and P.~Schaumont, ``{SIMON} says, break the area records
  for symmetric key block ciphers on fpgas,'' \emph{{IACR} Cryptol. ePrint
  Arch.}, p. 237, 2014. [Online]. Available:
  \url{http://eprint.iacr.org/2014/237}
\BIBentrySTDinterwordspacing

\bibitem{art12}
R.~Beaulieu, S.~Treatman-Clark, D.~Shors, B.~Weeks, J.~Smith, and L.~Wingers,
  ``The simon and speck lightweight block ciphers,'' in \emph{2015 52nd
  ACM/EDAC/IEEE Design Automation Conference (DAC)}, 2015, pp. 1--6.

\bibitem{art16}
\BIBentryALTinterwordspacing
P.~Ahir, M.~Mozaffari-Kermani, and R.~Azarderakhsh, ``Lightweight architectures
  for reliable and fault detection simon and speck cryptographic algorithms on
  fpga,'' \emph{ACM Trans. Embed. Comput. Syst.}, vol.~16, no.~4, may 2017.
  [Online]. Available: \url{https://doi.org/10.1145/3055514}
\BIBentrySTDinterwordspacing

\bibitem{Rashidi19}
\BIBentryALTinterwordspacing
B.~Rashidi, ``High-throughput and flexible {ASIC} implementations of {SIMON}
  and {SPECK} lightweight block ciphers,'' \emph{Int. J. Circuit Theory Appl.},
  vol.~47, no.~8, pp. 1254--1268, 2019. [Online]. Available:
  \url{https://doi.org/10.1002/cta.2645}
\BIBentrySTDinterwordspacing

\bibitem{Sheikhpour}
\BIBentryALTinterwordspacing
S.~Sheikhpour, M.~H. Sadi, and A.~Mahani, ``High-throughput configurable
  {SIMON} architecture for flexible security,'' \emph{Microelectron. J.}, vol.
  113, p. 105085, 2021. [Online]. Available:
  \url{https://doi.org/10.1016/j.mejo.2021.105085}
\BIBentrySTDinterwordspacing

\bibitem{Abed}
\BIBentryALTinterwordspacing
S.~Abed, R.~Jaffal, B.~J. Mohd, and M.~Alshayeji, ``{FPGA} modeling and
  optimization of a {SIMON} lightweight block cipher,'' \emph{Sensors},
  vol.~19, no.~4, p. 913, 2019. [Online]. Available:
  \url{https://doi.org/10.3390/s19040913}
\BIBentrySTDinterwordspacing

\end{thebibliography}

\section*{NOTICE}

\copyright 2022 IEEE. Personal use of this material is permitted. Permission from IEEE must be obtained for all other uses, in any current or future media, including reprinting/republishing  this material for advertising or promotional purposes, creating new collective works, for resale or redistribution to servers or lists, or reuse of any copyrighted component of this work in other works.
\end{document}